\documentclass[twocolumn,pre,superscriptaddress,nofootinbib]{revtex4-2}
\usepackage{amsfonts}
\usepackage{amsmath}
\usepackage{amssymb}
\usepackage{color}
\usepackage[toc,page]{appendix}
\usepackage{graphicx}
\usepackage{bm}
\usepackage{esint}
\usepackage{ulem}
\usepackage{xcolor}

\usepackage{soul}
\usepackage{cancel}
\usepackage{stmaryrd}

\begin{document}

\title{Bloch diode}

\date{\today}
\author{M.~Houzet}
\affiliation{Univ.~Grenoble Alpes, CEA, Grenoble INP, IRIG, Pheliqs, F-38000 Grenoble, France}
\author{T.~Vakhtel}
\affiliation{QuTech and Kavli Institute of Nanoscience, Delft University of Technology, Delft 2628 CJ, The Netherlands}
\author{J.~S.~Meyer}
\affiliation{Univ.~Grenoble Alpes, CEA, Grenoble INP, IRIG, Pheliqs, F-38000 Grenoble, France}
\begin{abstract}
In a SQUID tuned away from half-integer flux (in units of the superconducting flux quantum), the concurrence of multiple Josephson harmonics and an asymmetry between the junctions leads to the \textit{Josephson diode effect} — a nonreciprocal current-voltage characteristic manifested as an asymmetry of critical currents at opposite polarities. We predict a dual version of this effect in a gate-tunable Cooper pair transistor placed in series with a highly resistive environment. When tuned away from half-integer gate charge (in units of the Cooper pair charge) it shows an asymmetry of critical voltages at opposite polarities --  a dual diode effect we refer to as the \textit{Bloch diode effect}. It arises from an asymmetry in the dispersion of the transistor’s Bloch bands. A highly resistive environment can be realized with a Josephson junction array, suggesting that such a diode could be implemented using conventional superconducting quantum circuits.
\end{abstract}
\maketitle

The Josephson diode effect is characterized by asymmetric critical currents for opposite directions of the current flow across a Josephson junction between two superconducting leads~\cite{Nadeem2023}. The breaking of both inversion and time-reversal symmetries is necessary to generate an equilibrium current-phase relation that displays such critical current asymmetry. Out of equilibrium, this asymmetry then results in a nonreciprocal current-voltage characteristic as the junction is embedded in a dissipative environment. In particular, the Josephson diode may be used as a ratchet, a device able to rectify a signal with zero time-average~\cite{Zapata1996}. The interplay of magnetism and spin-orbit coupling provides a nontrivial contribution to the Josephson diode effect~\cite{Reynoso2012} and motivates the recent resurgence of interest in its study in Josephson junctions made with exotic materials~\cite{Nadeem2023}. Still, it has been known for a long time that Josephson diodes may also exist in the absence of spin-orbit coupling, provided that inversion symmetry is broken at a macroscopic, rather than microscopic level~\cite{Fulton1972}. In particular, they have been recently studied theoretically~\cite{Seoane2022,Fominov2022} and experimentally~\cite{Valentini2024,Leblanc2024} in asymmetric superconducting quantum-interference devices (SQUIDs) with more than one harmonic in the current-phase relation of the Josephson junctions that constitute them. 

Contrasting views about the role of Coulomb interaction on the nonreciprocal response of a Josephson junction have been presented. On the one hand, quantum fluctuations due to charging effects are detrimental to the Josephson diode effect described above~\cite{Scheidl2002,Hamamoto2019}. Indeed, the latter is associated with an equilibrium current-phase relation that breaks parity symmetry, $I_J(-\varphi)\neq -I_J(\varphi)$. Now let us add to the Josephson potential $U(\varphi)$, such that $I_J=(2e/\hbar)\partial_\varphi U$,  the standard Coulomb potential quadratic in the charge variable canonically conjugate to $\varphi$. Then, as is well known from quantum mechanics~\cite{Landau}, the Bloch wavefunctions $u_q$ of quasi-momentum $q$ for the resulting Hamiltonian remain symmetric under inversion of quasi-momentum, $u_q = u_{-q}$, despite the asymmetry of the Josephson potential, $U(-\varphi)\neq U(\varphi)$. Introducing a dissipative environment to address the current-voltage characteristic, one eventually finds that this absence of chirality in turn results in the suppression of any nonreciprocal response at low temperature as soon as the resistance of the environment exceeds the resistance quantum and the phase remains quantum-mechanically delocalized~\cite{Schmid1983}. On the other hand, nonreciprocity is ubiquitous in conventional semiconducting diodes, such as a $pn$ junction, where Onsager relations are bypassed by charge redistribution at finite bias. This observation questions the necessity of breaking time-reversal symmetry already in equilibrium, like in a Josephson diode, in order to get a nonreciprocal current-voltage characteristic. In fact, nonreciprocity was predicted in various phenomenological models of Josephson junctions preserving time-reversal symmetry, such that $I_J(-\varphi)= -I_J(\varphi)$, together with a non-quadratic electrostatic potential due to imperfect screening in semiconducting areas of the device~\cite{Hu2007,Misaki2021,Zhang2022}.

\begin{figure}
\includegraphics[width=.8\columnwidth]{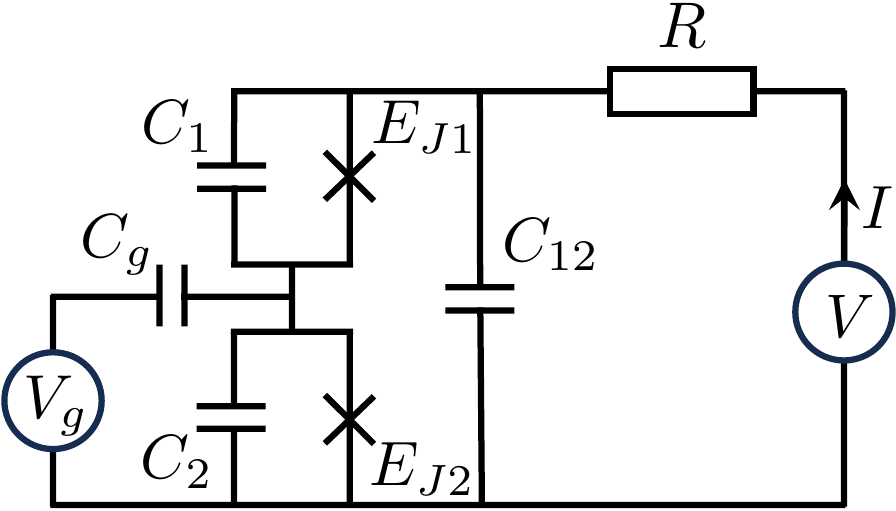}
\caption{\label{F:circuit}
The Bloch diode effect is predicted in a voltage-biased asymmetric Cooper pair transistor connected in series with a highly-resistive environment, $R\gg R_Q$.}
\end{figure} 

\begin{figure}
\includegraphics[width=\columnwidth]{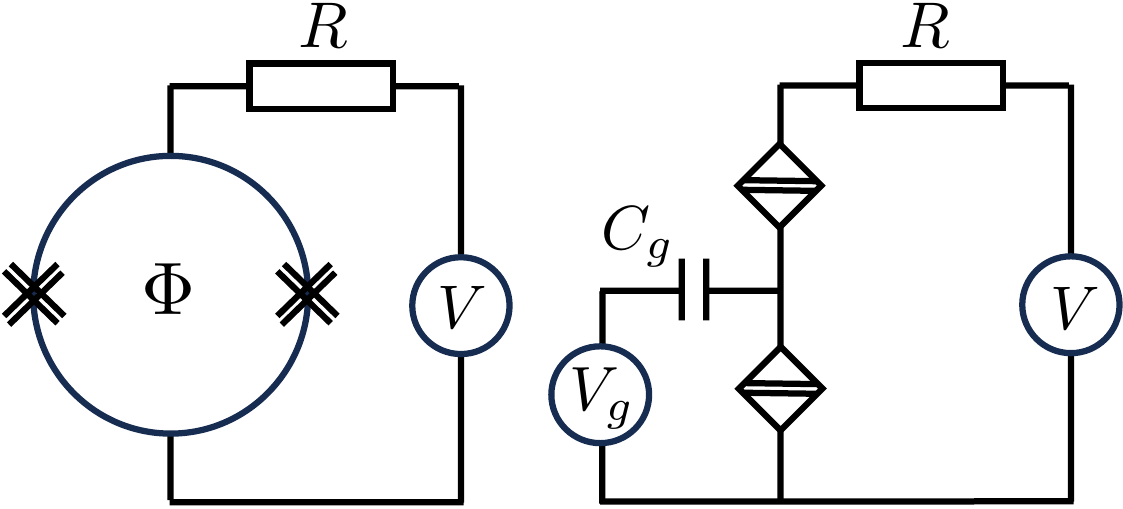}
{\caption{\label{F:dual}
Duality between the Josephson diode effect in a SQUID at $R\ll R_Q$ (left) and the Bloch diode effect in a Cooper pair transistor at $R\gg R_Q$ (right). The double cross represents a Josephson junction with multiple harmonics in its current-phase relation; the hatched diamond represents a quantum phase-slip junction with multiple harmonics in its voltage-charge relation. Time-reversal symmetry enforces $U(-\varphi,-\Phi)=U(\varphi,\Phi)$ for the Josephson potential of the SQUID, where $\Phi$ is the flux piercing the loop. Thus nonreciprocity is realized if $\Phi$ is tuned away from a half-integer flux (in units of the superconducting flux quantum) and the SQUID is asymmetric, such that $U(-\varphi,\Phi)\neq U(\varphi,\Phi)$. In the Cooper pair transistor, inversion symmetry enforces ${\cal E}_0(-{\cal N},-{\cal N}_g)={\cal E}_0({\cal N},{\cal N}_g)$ with ${\cal N}_g=-C_gV_g/2e$. Thus nonreciprocity is realized for two unequal phase-slip junctions if the gate charge is tuned away from a multiple of $e$, such that ${\cal E}_0(-{\cal N},{\cal N}_g)\neq{\cal E}_0({\cal N},{\cal N}_g)$. 
}}
\end{figure}

In this work we discuss a concrete Josephson circuit with broken inversion symmetry only, see Fig.~\ref{F:circuit}, which allows realizing a nonreciprocal current-voltage characteristic. We rely on the phase/charge duality predicted by Schmid~\cite{Schmid1983} to propose that a Cooper pair (or Bloch) transistor~\cite{Averin1991,Fulton1989,Geerligs1990,Tuominen1992} in series with a highly-resistive environment realizes a dual version of the resistively shunted SQUID with multiple Josephson harmonics{, see Fig.~\ref{F:dual}}. In the dual regime, the critical current asymmetry in the conventional Josephson diode is replaced by an asymmetry of critical voltages below which Coulomb blockade of Cooper pair tunneling takes place~\cite{Haviland1994,Zorin1999,Lotkhov2003,Watanabe2004,Corlevi2006}. In our proposal, the breaking of inversion symmetry originates from an asymmetry either in charging or Josephson energy between the two Josephson junctions that form the Cooper pair transistor. At the same time, our proposal only requires an array of conventional metallic tunnel junctions. The nonreciprocal dc current that flows through the device is accompanied by an ac current that manifests Bloch oscillations, dual to the ac Josephson effect~\cite{Likharev1985,Averin1985}. Therefore we suggest that the dual Josephson diode studied in this work should be dubbed the ``Bloch diode''. The recent observation of dual Shapiro steps induced by the mixing of microwaves with Bloch oscillations in voltage-biased single- and double-junction devices~\cite{Kaap2024,Kaap2024b,Shaikhaidarov2024}, which was enabled by their connection in series with a highly-resistive environment, is encouraging for the observation of the Bloch diode effect.

\vspace{.5cm}

At large environment resistance $R\gg R_Q$, where $R_Q=h/4e^2$ is the resistance quantum, the Cooper pair transistor in Fig.~\ref{F:circuit} behaves as a nonlinear capacitor~\cite{Sillanpaa2005,Duty2005}. The current-voltage characteristic of the circuit is then derived from the equation of motion
\begin{equation}
\label{eq:EOM}
V=RI+V_B({\cal N}), \qquad V_B({\cal N})=-\frac{1}{2e}\frac{\partial {\cal E}_0}{\partial {\cal N}},
\end{equation}
which expresses the bias voltage $V$ as the sum of the voltage drops at the resistor and at the effective nonlinear capacitor, respectively~\cite{Likharev1985,Averin1985,Averin1991}. Here, the current $I=-2e\dot{\cal N}$ is related with the charge $\cal N$ (in units of the Cooper pair charge) displaced across the resistor, and ${\cal E}_0({\cal N})$ is the ground-state energy of the Cooper pair transistor when it is electrostatically tunable. The (Bloch) voltage-charge relation $V_B({\cal N})$ plays a dual role to the (Josephson) current-phase relation $I_J(\varphi)$, while the relation between $I$ and $\dot{\cal  N}$ mirrors the second Josephson relation between $V$ and $\dot \varphi$. 

Current is blocked in a voltage range $V_{c-}<V<V_{c+}$ with critical voltages
\begin{equation}
\label{eq:Vc}
V_{c-}=\min_{\cal N}V_B({\cal N}),\quad
V_{c+}=\max_{\cal N}V_B({\cal N}).
\end{equation}
Generically $V_{c-}\neq -V_{c+}$ for an asymmetric device. This asymmetry then develops into a nonreciprocal current-voltage characteristic beyond the blocked region. Below we determine the amplitude of this asymmetry by evaluating the so-called ``dual diode efficiency''
\begin{equation}
\eta=\frac{V_{c+}-|V_{c-}|}{V_{c+}+|V_{c-}|}
\end{equation}
with $\eta_{\rm max}$ defined as the maximum of $|\eta|$. To find  ${\cal E}_0$ that appears in Eq.~\eqref{eq:EOM}, we introduce the Hamiltonian that describes an electrostatically controlled Cooper pair transistor,
\begin{eqnarray}
\label{eq:H0}
H&=&-E_{J1}\cos\varphi_1 -E_{J2}\cos\varphi_2+ E_{C1}(N_1-{\cal N})^2
\\
&&+ E_{C2}(N_2-{\cal N}-{\cal N}_g)^2
+ E_{C0}(N_2-N_1-{\cal N}_g)^2.
\nonumber
\end{eqnarray}
Here $\varphi_k$ and $-2eN_k$ are the canonically conjugate superconducting phase difference and Cooper pair charge that tunneled across the Josephson junction labelled by $k=1,2$, such that $[\varphi_k,N_{k'}]=i\delta_{k,k'}$. The two first terms in $H$ describe Cooper-pair tunneling characterized by Josephson energies $E_{J1}$ and $E_{J2}$. The last three terms involve external charges $\cal N$ and ${\cal N}_g=-C_gV_g/2e$ with gate capacitance $C_g$ and gate voltage $V_g$, as well as charging energies $E_{C1}=2e^2 {\tilde C}_2/C_{\rm eff}^2$, $ E_{C2}=2e^2C_1/C_{\rm eff}^2$, and $ E_{C0}=2e^2C_{12}/C_{\rm eff}^2$ with junction capacitances $C_1$ and $C_2$, cross-capacitance $C_{12}$, and $C_{\rm eff}^2=C_1{\tilde C}_2+C_{12}(C_1+{\tilde C}_2)$ with $\tilde C_2=C_2+C_g$. 
(From now on we replace $\tilde C_2$ with $C_2$ in all formulas assuming, e.g.,  $C_g\ll C_2$.) 

Within this model, the asymmetry needed for a finite diode efficiency requires $C_1\neq C_2$ or $E_{J1}\neq E_{J2}$. Furthermore, eigenfunctions $\Psi(\varphi_1,\varphi_2)$ of Eq.~\eqref{eq:H0} are $2\pi$-periodic in variables $\varphi_1,\varphi_2$. As a consequence, ${\cal E}_0$ is 1-periodic in variables ${\cal N},{\cal N}_g$ and ${\cal E}_0(-{\cal N},-{\cal N}_g)={\cal E}_0({\cal N},{\cal N}_g)$. We deduce that the diode efficiency is an odd and 1-periodic function of variable ${\cal N}_g$, which vanishes if ${\cal N}_g=0\, {\rm mod}\, \frac12$. Thus we restrict our study to the range $0<{\cal N}_g<\frac12$.

\vspace{.5cm}

We first consider the regime $C_{12}\ll C_1,C_2$ and ignore the last term in Eq.~\eqref{eq:H0}, such that the Hamiltonian becomes separable in variables $(\varphi_1,N_1)$ and $(\varphi_2,N_2)$. Then, 
\begin{equation}
\label{eq:SumE0}
{\cal E}_0({\cal N},{\cal N}_g)={\cal E}_1({\cal N})+{\cal E}_2({\cal N}+{\cal N}_g),
\end{equation} 
where ${\cal E}_k({\cal N})$ is the ground state energy of the Hamiltonian
\begin{equation}
\label{eq:1JJ}
H_k=-E_{Jk}\cos\varphi_k+E_{Ck}(N_k-{\cal N})^2.
\end{equation}
In particular,
\begin{equation}
\label{eq:EN}
{\cal E}_k{(\cal N})\approx\left\{
\begin{array}{ll}
\min_NE_{Ck}({\cal N}-N)^2,&E_{Jk}\ll E_{Ck},\\
 {\cal E}^0_k-\lambda_k\cos2\pi {\cal N},&E_{Jk}\gg E_{Ck},
\end{array} 
\right.
\end{equation}
with  ${\cal E}^0_k\approx -E_{Jk}+\frac12\hbar\omega_k$, quantum phase slip amplitude
\begin{equation}
\lambda_k=\frac{8}{\sqrt{\pi}}(2E_{Jk}^3E_{Ck})^{1/4}e^{-\sqrt{32 E_{Jk}/{E_{Ck}}}},
\end{equation}
and Josephson plasma frequency $\omega_{k}=\frac1\hbar \sqrt{2E_{Jk}E_{Ck}}$~\cite{Koch2007}.

A single junction is not sufficient to obtain a diode. Thus, if the contribution of one of the two junctions dominates over the other one, say ${\cal E}_0\approx {\cal E}_1({\cal N})$ for concreteness, then the diode efficiency vanishes. In that case, as the ratio $E_{J1}/E_{C1}$ increases, $V_{c+}=-V_{c-}$ decreases from $e/C_1$ to $\pi \lambda_1/e$. By contrast, when both junctions provide comparable contributions to ${\cal E}_0$, the diode efficiency is finite. In particular, in the Coulomb dominated regime, assuming $C_1<C_2$ for concreteness, we find
\begin{eqnarray}
\eta=\left\{\begin{array}{ll}
\frac{C_2-C_1}{C_1+C_2}\frac{{\cal N}_g}{1-{\cal N}_g},&0<{\cal N}_g<\frac{C_1}{C_1+C_2},\\
\frac{C_{1}}{C_{2}}(1-{2\cal N}_g),&\frac{C_1}{C_1+C_2}<{\cal N}_g<\frac12.
\end{array}\right.
\end{eqnarray}
The maximal diode efficiency, $\eta_{\rm max}=(\sqrt{2}-1)^2\simeq0.17$, is reached at $C_1/C_2=\sqrt{2}-1$ and ${\cal N}_g=1-1/\sqrt{2}$. 

\begin{figure}
\includegraphics[width=.8\columnwidth]{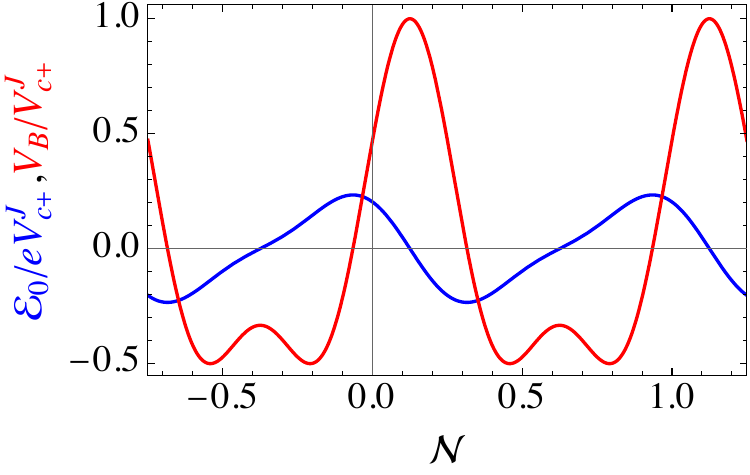}
\caption{\label{F:2}
Ratchet potential (blue) and voltage-charge relation (red) associated with a Bloch diode with efficiency $\eta=1/3$ { in the Josephson-dominated regime}, cf.~Eq.~\eqref{eq:ratchet}.} 
\end{figure}

In the Josephson dominated regime, by inserting the second line of Eq.~\eqref{eq:EN} into Eq.~\eqref{eq:SumE0}, we find
a modulation of the critical voltages,
\begin{equation}
\label{eq:Vc0}
V_{c\pm}^{J0}=\pm \frac\pi e\sqrt{(\lambda_1-\lambda_2)^2+4\lambda_1\lambda_2\cos^2\pi {\cal N}_g}
\end{equation}
(without diode effect) due to Aharonov-Casher (AC) interference~\cite{Ivanov2001,Friedman2002}, which is dual to the flux modulation of a SQUID. When $\lambda_1\approx \lambda_2$ and ${\cal N}_g\approx\frac12$, the blocked region is suppressed. In that case, it is important to keep the next-to-leading order term~\cite{ZinnJustin1981,Vakhtel2024} in the second line of Eq.~\eqref{eq:EN} (see Appendix~\ref{sec:doubleQPS} for the derivation of its amplitude),
\begin{equation}
\label{eq:doubleQPS}
{\cal E}_k{(\cal N})\approx {\cal E}^0_k-\lambda_k\cos2\pi {\cal N}
+\frac{\lambda^2_k}{2\hbar\omega_{k}}\ln\left(\frac{2^4 e^\gamma\hbar \omega_{k}}{E_{Ck}}\right)\cos4\pi{\cal N},
\end{equation}
where $\gamma$ is the Euler constant, to compute the critical voltage.
(Note that the amplitude of that term, also called ``double quantum phase slip" amplitude, was recently measured in a fluxonium qubit~\cite{Ardati2024}.) The dual diode effect then results from different AC-induced phase shifts in the first and second harmonics of the voltage-gate charge relation, similar to the case of a SQUID with higher harmonic content~\cite{Seoane2022,Fominov2022}. Following the analysis of Ref.~\cite{Volkov2024} for the conventional Josephson diode effect, we find that the maximal dual diode efficiency, $\eta_{\rm max}=1/3$, is reached when
\begin{equation}
\label{eq:maxeta}
\frac{|\lambda_1-\lambda_2|}{\lambda_1}
=\frac{\sqrt{2}\lambda_1}{\hbar \omega_{1}}\ln\left(\frac{E_{J1}}{E_{C1}}\right)
=2\pi |{\cal N}_g-\frac12|\ll 1,
\end{equation}
corresponding to
\begin{equation}
\label{eq:ratchet}
{\cal E}_0({\cal N})={\sqrt{2}|\lambda_1-\lambda_2|} \left[\cos 2\pi ({\cal N}+\frac18)+\frac14\cos 4\pi {\cal N}\right].
\end{equation}
(Here we also assumed $\omega_1\approx \omega_2$ for simplicity.) The charge dispersion of ${\cal E}_0$ is illustrated in Fig.~\ref{F:2}; it forms a ratchet potential. The asymmetry of critical voltages{, $V^J_{c+}=-2V^J_{c-}=(3\pi/\sqrt{2})|\lambda_1-\lambda_2|/e$,} is also seen in the voltage-charge relation. 

The diode efficiency in an hybrid situation where one junction is in the Josephson dominated regime, and the other one in the Coulomb dominated regime is addressed in Appendix~\ref{sec:hybrid}.

\vspace{.5cm}

Let us now consider the case when the Hamiltonian~\eqref{eq:H0} is not separable. We first consider the case when $E_{C0}$ dominates over all other energy scales, $E_{C0}\approx 2e^2/(C_1+C_2)\gg E_{C1},E_{C2}$ at $C_{12}\gg C_1,C_2$. Then charge in the central island is strongly pinned, and transport can only occur in the vicinity of a degeneracy point between two charge states. For concreteness, we consider $|{\cal N}_g-\frac12|\ll 1$ and project the Hamiltonian \eqref{eq:H0} onto the charge basis with $N_2-N_1=0,1$. Then,
\begin{eqnarray}
H&=&E_{C}\left(N-{\cal N}+\frac{\delta E_{C}}{4E_{C}}\sigma_z\right)^2-E_{C0}\left(\frac12-{\cal N}_g\right)\sigma_z
\nonumber
\\
&&-E_{J}\cos\frac\varphi 2\sigma_x+\delta E_{J}\sin\frac\varphi 2\sigma_y
\label{eq:H1}
\end{eqnarray}
(up to irrelevant constants). Here we introduced Pauli matrices in the basis of the two charge states of the island, and canonically conjugate variables $\varphi=\varphi_1+\varphi_2$ and $N=(N_1+N_2)/2$. Furthermore, $E_C=E_{C1}+E_{C2}$, $\delta E_C=E_{C1}-E_{C2}$, $E_J=(E_{J1}+E_{J2})/2$, and $\delta E_J=(E_{J1}-E_{J2})/2$. Charge conservation imposes the twisted boundary condition $\Psi(\varphi+2\pi)=\sigma_z\Psi(\varphi)$ on eigenstates of Eq.~\eqref{eq:H1} \cite{Vakhtel2023}. As discussed recently~\cite{Mert2024}, Hamiltonian~\eqref{eq:H1} has the same form as the Hamiltonian for a discrete resonant level coupled to superconducting leads at weak Coulomb interaction~\cite{Kurilovich2021, Vakhtel2023,Gungordu2025}.

\begin{figure}
\includegraphics[width=.8\columnwidth]{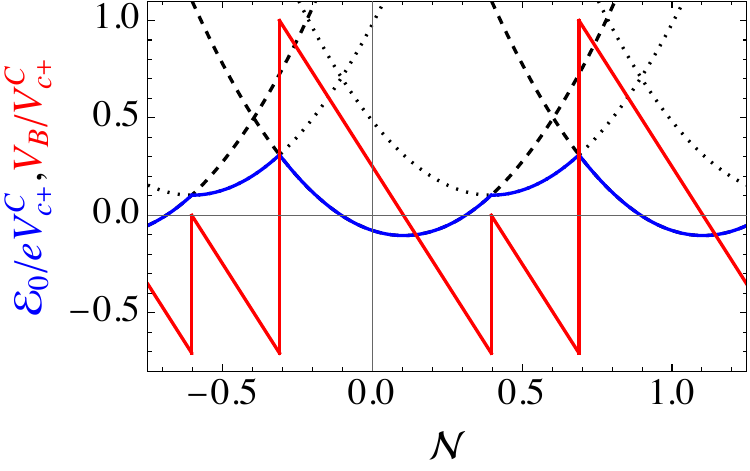}
{\caption{\label{F:4}
Ratchet potential (blue) and voltage-charge relation (red) associated with a Bloch diode with efficiency $\eta=(\sqrt{2}-1)^2$ in the Coulomb-dominated regime. Dashed and dotted lines are metastable energies at fixed charge and $\sigma_z=\mp 1$, respectively, cf.~Eq.~\eqref
{eq:E0-CB}. The gaps of the order $E_J\ll E_C$ when charging energies cross each other, and associated smearing of the charge-voltage relation, are not shown.}}
\end{figure}

In the Coulomb dominated regime, $\sigma_z=\pm$1 and ${\cal E}_0=\min({\cal E}_+,{\cal E}_-)$ with 
\begin{equation}
\label{eq:E0-CB}
\frac{{\cal E}_\pm}{E_C}=\min_{n\in\mathbb{Z}}\left(n-{\cal N}\mp\frac\delta4+\frac{1\mp 1}4\right)^2\pm \varepsilon,
\end{equation}
$\delta=\delta E_C/E_{C}$, and $\varepsilon= E_{C0}\left(\frac12-{\cal N}_g\right)/E_C$. As before we consider $C_1<C_2$ corresponding to $\delta>0$, where we find the diode efficiency
\begin{equation}
\eta=\left\{\begin{array}{ll}
\frac{8\varepsilon}{(1{+}\delta)^2},&0<\varepsilon<\frac\delta{8}(1- \delta^2),\\
\delta\frac{1-\delta^2-8\varepsilon}{1-\delta^2+8\varepsilon},&\frac\delta{8}(1- \delta^2)<\varepsilon<\frac1{8}(1- \delta^2).
\end{array}
\right.\label{eq-diode_C12}
\end{equation}
At $\varepsilon>\frac1{8}(1- \delta^2)$, the energies ${\cal E}_\pm$ do not cross and the diode efficiency vanishes. Eq.~\eqref{eq-diode_C12}
yields the same diode efficiency as in the separable case, $\eta_{\rm max}=(\sqrt{2}-1)^2$, which is realized at $\delta=\sqrt{2}-1$ and $\varepsilon=(\sqrt{2}-1)^2/4$. In fact, the result for $\eta_{\rm max}=(\sqrt{2}-1)^2$ holds in the Coulomb dominated regime even without assuming $C_{12}\gg C_1,C_2$. In that case, $\varepsilon=C_{\rm eff}^2/(C_1+C_2)^2(\frac12-{\cal N}_g)$. {The $\cal N$-dependence of ${\cal E}_0$ and charge-voltage relation at maximal diode efficiency, with $V^C_{c+}=-\sqrt{2}V^C_{c-}=(\sqrt{2}-1)E_C/e$, are shown in Fig.~\ref{F:4}.}

\begin{figure}
\includegraphics[width=.8\columnwidth]{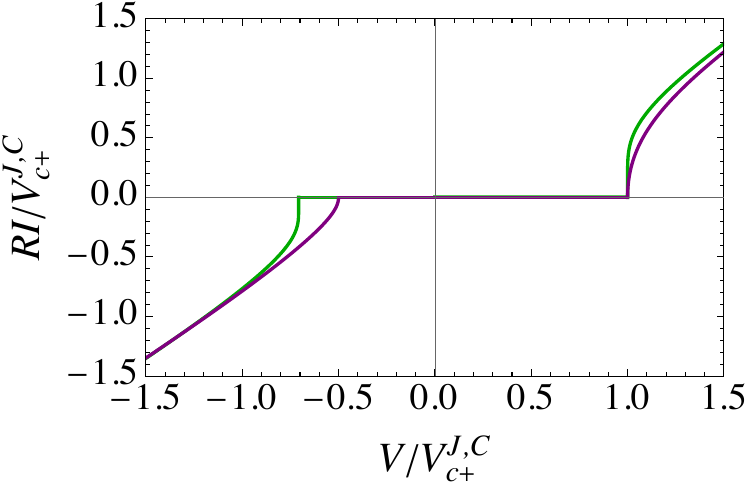}
{\caption{\label{F:3}
Nonreciprocal current-voltage characteristics of the Bloch diode described by the ratchet potential of Fig.~\ref{F:2} (purple) and Fig.~\ref{F:4} (green).}}
\end{figure}

In the Josephson dominated regime, the WKB approximation was used to calculate the two first harmonics in the $\cal N$-dispersion of ${\cal E}_0$ at $C_{12}\gg C_1=C_2$, see Eqs.~(19) and (25) in Ref.~\cite{Vakhtel2023}, where the difference between AC shifts in the first and second harmonics in $2\pi{\cal N}$ is independently tunable with ${\cal N}_g$. As in the separable case, we then expect $\eta_{\rm max}=1/3$. In fact, the result for $\eta_{\rm max}=1/3$ carries on at any ratio between capacitances $C_{12}, C_1,C_2$ in the Josephson dominated regime, where a multidimensional WKB analysis also shows that the $\cal N$-dispersion of  ${\cal E}_0$ is dominated by the two first harmonics with an independently tunable difference between AC shifts \cite{Kaap2024b,Semenov2024}.

The nonreciprocity in the current-voltage characteristics of a Bloch diode is illustrated in Fig.~\ref{F:3} by solving Eq.~\eqref{eq:EOM} for the Bloch diode { with maximal efficiency $\eta=\frac13$ in the Josephson-dominated regime and $\eta=(\sqrt{2}-1)^2$ in the Coulomb-dominated regime. Near voltage thresholds $V_{c\pm}$, the current displays inverse square-root and logarithmic singularities, respectively. Note that this theory neglects interband transitions. In the Josephson-dominated regime, this is guaranteed over a large voltage range by the gap between Bloch bands (of the order of the Josephson plasma frequency); in the Coulomb dominated regime, the condition for neglecting Landau-Zener tunneling to the upper bands is $e|V|\ll (E_J^2/E_C)(R/R_Q)$~\cite{Averin1985}.} 
Thermal and quantum fluctuations are expected to smooth out the nonreciprocal response at $k_BT\sim eV_{c\pm}$ and $R\sim R_Q$, respectively.

\vspace{.5cm}

The Bloch diode that we studied is related with the so-called ``bifluxon" qubit~\cite{Kalashnikov2020}, where a Cooper pair transistor is closed by an inductance to form a flux-tunable loop. The bifluxon provides an example of a ``few-body" qubit with more than one degree of freedom in the Hamiltonian that describes it, which allows suppressing relaxation due to charge or flux noise, and dephasing due to flux noise. That protection is expected at ${\cal N}_g=\frac12$ provided that the device is fully symmetric ($E_{J1}=E_{J2}$ and $C_1=C_2$){~\cite{Comment2002}}. By contrast, the Bloch diode that we predict is precisely due to the device asymmetry. Conversely, measuring the (disappearance of the) Bloch diode effect may help to tune a device to the most symmetric point, such that at ${\cal N}_g=1/2$ only the double quantum phase slip process ($\propto \cos 4\pi {\cal N}$) remains. In practice, however, if the timescale of quasiparticle poisoning is not much larger than the measurement timescale, poisoning events can shift either $ \mathcal N$ or $ \mathcal N_g $ by $ 1/2$. Due to the presence of more than one harmonic, this will generally affect the critical voltages, eventually reducing the Bloch diode efficiency.

Finally, we note that transmission lines formed of a Josephson junction array can be used to realize a highly resistive ohmic environment~\cite{Watanabe2004,Corlevi2006,Masluk2012}. We thus envision that the Bloch diode effect predicted in our work could be observed in a purely non-dissipative superconducting quantum circuit. 

TV acknowledges insightful discussions with P.~Kurilovich, V.~Kurilovich, and S.~Bosco.
MH and JSM acknowledge funding from the Plan France 2030 through the project NISQ2LSQ ANR-22-PETQ-0006 and FERBO ANR-23-CE47-0004.

\appendix

\section{``Double quantum phase slip'' amplitude}
\label{sec:doubleQPS}

Here we derive the exponentially small dispersion of the ground state energy of a Cooper pair box, which is described by Hamiltonian \eqref{eq:1JJ}, in the Josephson-dominated regime up to the second harmonic in gate charge $\cal N$, see Eq.~\eqref{eq:doubleQPS}.

For this we note that, up to a gauge transformation, the eigenenergies of a Cooper pair box are defined by the following eigenproblem:
\begin{align}
\left[-E_C \frac{d^2}{d\varphi^2}-E_J\cos\varphi\right]\psi(\varphi)=E\psi(\varphi),
\label{eq:Schroedinger}
\end{align}
where the wavefunction $\psi(\varphi)$ must satisfy the twisted periodic boundary condition
\begin{align}
\psi(\varphi+2\pi)=e^{-i2\pi \mathcal{N}}\psi(\varphi).
\label{eq:BC}
\end{align}
Here  for simplicity we omit the index $k=1,2$ used in the main text. Our strategy to obtain Eq.~\eqref{eq:doubleQPS} is to solve perturbatively a transcendental equation solved by $E({\cal N})$. To find that equation, we determine the most general forms of the solutions of Eq.~\eqref{eq:Schroedinger} for phases in the vicinity of multiples of $2\pi$, where $\psi(\varphi)$ has most of its support at $E_J\gg E_C$, as well as in the classically forbidden regions, and then match them in a common range in $\varphi$, where both these solutions are defined. 

Let us first consider the vicinity of $\varphi=0$. We may use an harmonic approximation of the Josephson potential, $\cos\varphi\approx 1-\varphi^2/2$, and rescaled variables $\varphi=x\sqrt[4]{{E_{C}}/{2E_{J}}}$ and $E=-E_J+\hbar \omega(p+1/2)$ with plasma frequency $\hbar \omega=\sqrt{2E_JE_C}$, to find that Eq.~\eqref{eq:Schroedinger} reduces to
\begin{align}
\left(\frac{d^2}{dx^2}+p+\frac12-\frac{x^2}4\right)\psi(x)=0.
\end{align}
Its most general solution is expressed in terms of parabolic cylinder functions $D_p(x)$,
\begin{align}
\label{eq:QHO}
\psi(x)=A_{1}D_{p}(x)+A_{2}D_{p}(-x),
\end{align}
where $A_1$ and $A_2$ are two constants to be determined. The asymptotes of Eq.~\eqref{eq:QHO} at $x\gg 1$ and $-x\gg 1$ are
\begin{align}
\label{eq:asymptote1}
\psi(x)\approx\left(A_{1}+A_{2}e^{-i\pi p}\right)x^{p}e^{-{x^{2}}/{4}}-A_{2}\frac{\sqrt{2\pi}}{\Gamma(-p)}\frac{e^{{x^{2}}/{4}}}{x^{p+1}}\end{align}
and
\begin{align}
\label{eq:asymptote2}
\psi(x)\approx\left(A_{1}e^{i\pi p}+A_{2}\right)(-x)^{p}e^{-{x^{2}}/{4}}-A_{1}\frac{\sqrt{2\pi}}{\Gamma(-p)}\frac{e^{{x^{2}}/{4}}}{(-x)^{p+1}},
\end{align}
respectively. Note that the condition $|x|\gg 1$ is still compatible with $|\varphi|\ll 1$ used in the harmonic approximation for the Josephson potential at $E_J\gg E_C$. For $\psi(x)$ to remain localized near $\varphi=0$, the last term in Eqs.~\eqref{eq:asymptote1} and \eqref{eq:asymptote2} should be suppressed. Given the properties of the Euler function $\Gamma(-p)$, this requires that $p$ should be close to a positive integer $m$, in agreement with the expectation for the eigenenergies of a quantum harmonic oscillator,
\begin{align}
\label{eq:QHO2}
E_m=-E_J+\hbar\omega\left(m+\frac12\right)-\frac{E_C}{8}\left(m^2+m+\frac12\right).
\end{align}
Here we dispense with the derivation of a small, $\cal N$-independent anharmonic correction to the eigenenergies captured by the last term of Eq.~\eqref{eq:QHO2}~\cite{Koch2007}, which can obtained by expanding $\cos\varphi$ up to quartic order in $\varphi$.

Equivalent solutions with the same eigenspectrum can be found by expanding the Josephson potential near $\varphi=2\pi n$ with any integer $n$. It is their hybridization that results in the formation of Bloch bands. To proceed further, we then express the most general form of the wavefunction in an interval between two minima of the Josephson potential, say for $\varphi$ such that $\varphi\gg \sqrt[4]{{E_{C}}/{E_{J}}}$ and $2\pi-\varphi\gg \sqrt[4]{{E_{C}}/{2E_{J}}}$, using 
Wentzel-Kramers-Brillouin approximation in the classically forbidden regions,
\begin{align}
\label{eq:WKB}
\psi(\varphi) \approx
\frac{B_{1}}{\sqrt{\kappa(\varphi)}} e^{-\int_\pi^\varphi d\varphi' \kappa(\varphi')}+ \frac{B_{2}}{\sqrt{\kappa(\varphi)}} e^{\int_\pi^\varphi d\varphi' \kappa(\varphi')}.
\end{align}
Here $\kappa(\varphi)=\sqrt{(-E_J\cos\varphi-E)/E_C}$, and $B_1$ and $B_2$ are two other constants to be determined. (The choice of $\pi$ as a bound in the integrals that appear in Eq.~\eqref{eq:WKB} was made for convenience; another choice could be absorbed in $B_1$ and $B_2$.) As $\varphi$ remains away from the classical turning points (where the argument in $\kappa(\varphi)$ would vanish), we may approximate
\begin{align}
\label{eq:WKB-kappa}
\kappa(\varphi)\approx\sqrt{\frac{2E_J}{E_C}}\sin(\varphi/2)-\frac{p+\frac12}{2\sin(\varphi/2)},
\\
\int_\pi^\varphi d\varphi'\kappa(\varphi')\approx-\sqrt{\frac{8E_J}{E_C}}\cos\frac\varphi 2-\left(p+\frac12\right)\ln \tan\frac\varphi4.
\end{align}
As a result, Eq.~\eqref{eq:WKB} further simplifies to
\begin{align}
\psi(\varphi) \approx
\frac{B_1}{\sqrt{\sin(\varphi/2)}}\left(\tan\frac\varphi4\right)^{p+\frac12} e^{\sqrt{\frac{8E_J}{E_C}}\cos(\varphi/2)}\quad\nonumber\\
+\frac{B_2}{\sqrt{\sin(\varphi/2)}}\left(\tan\frac\varphi4\right)^{-p-\frac12} e^{-\sqrt{\frac{8E_J}{E_C}}\cos(\varphi/2)},
\label{eq:WKB2}
\end{align}
up to a common constant absorbed in $B_1,B_2$.

In particular, the asymptote of Eq.~\eqref{eq:WKB2} at $\sqrt[4]{E_C/E_J}\ll \varphi\ll 1$ has identical parametric dependence on $\varphi\propto x$ as in Eq.~\eqref{eq:asymptote1}. This allows us identifying two relations between coefficients $A_1,A_2,B_1,B_2$:
\begin{align}
A_{1}+A_{2}e^{-i\pi p}=\frac{B_1}{\sqrt{2}}\left(\frac{E_C}{2^9E_J}\right)^{ p/4}e^{\sqrt{8E_J/E_C}}, 
\label{eq:cond1} \\ 
A_{2}\frac{\sqrt{2\pi}}{\Gamma(-p)}=-\frac{B_2}{\sqrt{2}}\left(\frac{E_C}{2^9E_J}\right)^{-(p+1)/4}e^{-\sqrt{8E_J/E_C}},
\end{align}
Similarly, Eq.~\eqref{eq:WKB2} yields an asymptote at $\sqrt[4]{{E_{C}}/{2E_{J}}}\ll 2\pi-\varphi\ll1$, which can be put in correspondance with Eq.~\eqref{eq:asymptote2} using the periodic boundary condition given by Eq.~\eqref{eq:BC}. This yields two additional relations:
\begin{align}
A_{1}e^{i\pi p}+A_{2}=\frac{B_2}{\sqrt{2}}\left(\frac{E_C}{2^9E_J}\right)^{ p/4}e^{-\sqrt{8E_J/E_C}}e^{-2i\pi{\cal N}}, \\ 
A_{1}\frac{\sqrt{2\pi}}{\Gamma(-p)}=-\frac{B_1}{\sqrt{2}}\left(\frac{E_C}{2^9E_J}\right)^{-(p+1)/4}e^{\sqrt{8E_J/E_C}}e^{-2i\pi{\cal N}}.
\label{eq:cond4}
\end{align}
The linear system of equations \eqref{eq:cond1}-\eqref{eq:cond4} yields the sought-after transcendental equation solved by $\delta E_m\equiv E-E_m=\hbar\omega (p-m)$ with $p$ close to integer $m$:
\begin{align}
\label{eq:trans}
2\cos2\pi{\cal N}=\frac{\sqrt{2\pi}}{\Gamma(-p)}\left(\frac{E_C}{2^9E_J}\right)^{ p/2+1/4}e^{\sqrt{32E_J/E_C}}.
\end{align}
Using $1/\Gamma(-p)\approx(-1)^{m+1}m![(p-m)+\psi(m+1)(p-m)^2]$ at $|p-m|\ll 1$, where $\psi(m)$ is the digamma function, we expand Eq.~\eqref{eq:trans} to find
\begin{align}
\label{eq:sol}
\delta E_m=-\lambda_m\cos2\pi{\cal N}+\frac{\lambda_m^2}{2\hbar\omega}\ln\left(\frac{2^4\hbar \omega}{e^{\psi(m+1)}E_C}\right)\cos^22\pi{\cal N}.
\end{align}
with
\begin{align}
\frac{\lambda_m}{\hbar \omega}=\frac{(-1)^m}{m!}\sqrt{\frac 2\pi}\left(\frac{2^9E_J}{E_C}\right)^{m/2+1/4}e^{-\sqrt{32 E_J/E_C}}.
\end{align}
At $m=0$ with $\psi(1)=-\gamma$ where $\gamma$ is Euler's constant, Eq.~\eqref{eq:sol} reproduces Eq.~\eqref{eq:doubleQPS}.

\section{Diode efficiency in an hybrid regime}
\label{sec:hybrid}

Here we provide the diode efficiency in an hybrid regime where one of the junctions is in the Coulomb dominated regime (say $E_{J1}\ll E_{C1}$), and the other one is in the Josephson dominated regime (say $E_{J2}\gg E_{C2}$). If $y\equiv e^2/(\pi^2C_1\lambda_2)>1$,
\begin{equation}
\frac{eV_{c\pm}^{{>}}}{\pi \lambda_2}=\pm \pi y+\sin 2\pi {\cal N}_g, 
\end{equation}
yielding maximal diode efficiency $\eta_{\rm max}=1/\pi\approx0.32$ at ${\cal N}_g=\frac14$ and $y\to 1$.
If $y<1$, $V_{c+}=V_{c+}^{{<}}$ and $V_{c-}=\min(V^>_{c-},V^<_{c-})$ with
\begin{eqnarray}
\frac{eV_{c\pm}^{{<}}}{\pi \lambda_2}= \pm\left[\sqrt{1-y^2}+y\left(\pi - \arccos y\right)\right]+2\pi y{\cal N}_g,
\end{eqnarray}
yielding maximal diode efficiency $\eta_{\rm max}\approx 0.60$ at $y\approx 0.43$ and ${\cal N}_g\approx 0.39$.

\end{document}